# Spin wave and spin flip in hexagonal LuMnO$_3$ single crystal

Xiang-Bai Chen[1,*], Peng-Cheng Guo[1], In-Sang Yang[2,*], Xueyun Wang[3], Sang-Wook Cheong[3]

[1] School of Science and Laboratory of Optical Information Technology, Wuhan Institute of Technology, Wuhan 430205, China
[2] Department of Physics and Division of Nano-Sciences, Ewha Womans University, Seoul 120-750, Korea
[3] Rutgers Center for Emergent Materials and Department of Physics and Astronomy, Rutgers University, Piscataway, New Jersey 08854, USA

**Abstract**

Manipulate and control of spin wave and spin flip are crucial for future developments of magnonic and spintronic devices. We present that the spin wave in hexagonal LuMnO$_3$ single crystal can be selectively excited with laser polarization perpendicular to the *c*-axis of hexagonal LuMnO$_3$ and photon energy ~ 1.8 eV. The selective excitation of spin wave also suggests that the spin flip can be selectively controlled in hexagonal manganites. In addition, the physical origin of spin wave correlated with spin flip in hexagonal manganites is discussed.

[*] Corresponding authors: xchen@wit.edu.cn, yang@ewha.ac.kr

Today's electronic devices operate on electron charge. Basing on electron spin, a new generation of smaller, faster, and more efficient spintronic devices could be developed [1-5]. Recently, the idea of magnonic devices, in which spin waves (magnons) are carrying information, has received increasing attention in the magnetics community [6-8]. When developing spintronic and magnonic devices, it is crucial to control spin flip and spin wave. Therefore, understand the mechanisms of spin wave and spin flip in magnetic materials, and thus search for convenient method to control spin wave and spin flip would be of great importance for future developments of magnonic and spintronic devices.

Hexagonal manganites $R$MnO$_3$ ($R$ = rare earth) belong to the class of multiferroic oxides, which possess coexisting magnetic and ferroelectric phases, with cross-correlation effects between magnetic and electric degrees of freedom [9-12]. As such, they could potentially be used in applications of electric controlled magnonic and spintronic devices. In this study, we present interesting selective excitation of spin wave in hexagonal LuMnO$_3$ single crystal, and suggest the mechanism of selective control of spin flip. Also, the physical origin of spin wave correlated with spin flip in hexagonal manganites is discussed.

Hexagonal LuMnO$_3$ single crystal was grown using the traveling floating zone method. The characterizations of magnetization, resistivity, and x-ray powder diffraction have shown super crystalline quality of the single crystal sample. Polarized Raman scattering spectra were obtained in backscattering configuration with a Jobin-Yvon LabRam spectrometer. The single crystal sample was mounted in a helium closed cycle cryostat and the sample temperature was cooled to 20 K.

Figure 1 shows the Raman scattering spectra of hexagonal LuMnO$_3$ single crystal obtained at 20 K under three different polarization configuration conditions. The narrow peaks ~ 695 cm$^{-1}$ and 650 cm$^{-1}$ can be assigned to A$_1$ and E$_1$ phonon mode, respectively. The broad band ~ 815 cm$^{-1}$ would be originated from spin wave (magnon) scattering [13-15]. An interesting feature of Fig. 1 is that the spin wave at ~ 815 cm$^{-1}$ shows unique polarization selection rule: the spin wave can be excited under z(yx)z configuration, but not detectable under x(zy)x and y(zx)y configurations; i.e., the spin wave can be selectively excited with laser polarization parallel to the a-b plane of hexagonal manganites. Below Néel temperature, the spins of the Mn moments are aligned in the basal a-b plane of hexagonal manganites [16-17]. This may suggest that to excite spin wave scattering, laser polarization should be parallel to the plane of spin alignment.

In addition to the unique polarization selection rule, the spin wave of hexagonal manganites also exhibits unique resonance effect. Figure 2 shows the Raman scattering spectra of hexagonal LuMnO$_3$ single crystal obtained at 20 K under z(yx)z configuration with excitations of 633 nm red laser and 514 nm green laser. As can be seen in Fig. 2, the spin wave can be excited with 633 nm red laser, but not detectable with 514 nm green laser. This is consistent with our previous resonance study of spin wave in hexagonal HoMnO$_3$ thin film, which showed that the spin wave can be excited with 647 and 671 nm red lasers, but not detectable with 532 nm green laser (inset of Fig. 2) [13]. There are two main optical transitions in hexagonal manganites: a sharp peak at ~ 1.8 eV with linewidth ~ 0.2 eV, and a broad band centered at ~ 5 eV with linewidth ~ 2 eV [18-19]. The unique resonance effect of spin wave would be correlated with the narrow optical transition ~ 1.8 eV in hexagonal manganites.

The studies of electronic structure of hexagonal manganites showed that the optical transition ~ 1.8 eV is correlated with the on-site Mn *d-d* transitions [18-19] (transitions between Mn *d* orbitals hybridized with oxygen orbitals, since pure *d-d* transitions are not allowed). The on-site Mn *d-d* optical transitions are allowed with $E\perp c$ polarization of light; while in the case of $E\parallel c$ polarization of incident light, the optical matrix elements for the on-site Mn *d-d* transitions are zero [18]. Therefore, the unique polarization selection rule of the spin wave in hexagonal manganites would be also correlated with the on-site Mn *d-d* transitions.

Figure 1 and 2 indicated that the excitation of spin wave in hexagonal manganites can be selectively controlled with on-site Mn *d-d* transition. This selective excitation of spin wave in hexagonal manganites would be very helpful for their future applications in magnonic devices. In addition, spin wave is originated by excitation of spin flip. The selective excitation of spin wave in hexagonal manganites would suggest that spin flip in hexagonal manganites could also be selectively controlled, which is crucial for developing spintronic devices. Furthermore, it would be important to understand what kind of spin flip is correlated with the observed spin wave, i.e., the physical origin of spin wave. Below we will suggest the mechanism of spin flip and discuss the physical origin of spin wave in hexagonal manganites.

To understand the spin flip in hexagonal manganites, the knowledge of Mn *d-d* transition below Néel temperature would be helpful. The temperature dependent study of Mn *d-d* transition had shown that the transition has a blueshift of ~ 0.1 eV with cooling in the antiferromagnetic state [18-19]. Furthermore, Souchkov et al [18] suggested that this ~ 0.1 eV magnetic originated blueshift of the Mn *d-d* transition can be attributed to the differences in exchange interaction in the excited and ground states of a given Mn ion. The spin wave scattering in hexagonal manganite is observed ~ 815 cm$^{-1}$, i.e., has an energy of ~ 0.1 eV. This is in good agreement with the blueshift of the Mn *d-d* transition. The Heisenberg Hamiltonian model (to be discussed in the later part of this paper) indicated that the spin wave scattering at ~ 815 cm$^{-1}$ would be correlated with 4-spin-flip. Therefore, the ~ 0.1 eV blue shift of Mn *d-d* transition in the antiferromagnetic state would be correlated with 4-spin-flip.

The above results and discussions suggest that a possible process of spin flip, spin relaxing, and spin wave scattering in hexagonal manganites can be schematically represented as in Fig. 3. A microscopic picture of the process would be as following: when one Mn$^{3+}$ ion is resonantly excited, spin flip occurs, then during the spin relaxing, the spin flipped Mn$^{3+}$ ion interacts with the neighboring Mn$^{3+}$ ions, which changes only the direction of neighboring spin vectors, and thus forming spin wave propagating in hexagonal manganites. Our proposed model suggests that in ground state of Mn *d* orbital with S = 2, spin flip is quite difficult, i.e., ground state of Mn *d* orbital with S = − 2 is difficult to achieve. The spin flip can be much more easily achieved with resonant excitation of Mn *d-d* transition, i.e., excited state of Mn *d* orbital with S = − 2 can be much more easily achieved. Thus the experimental observed spin wave scattering would be originated from scattering of excited electronic states. In addition, Fig. 3 suggests that when resonantly pumped to excited states, spin wave scattering can also be observed with non-resonance probe source. To verify the proposed model in Fig. 3, extensive experimental and theoretical studies would be needed.

In addition the interesting selective excitation property of spin wave in hexagonal manganites, Fig. 1 and 2 also showed an abnormal observation. Due to the superb crystalline quality, the phonon mode of hexagonal LuMnO$_3$ single crystal did show narrower linewidth than that of hexagonal HoMnO$_3$ thin film. However, the spin wave of hexagonal LuMnO$_3$ single crystal showed significantly broader linewidth than that of hexagonal HoMnO$_3$ thin film. In addition, the spin wave of hexagonal LuMnO$_3$ single crystal is much more asymmetric than that of hexagonal HoMnO$_3$ thin film, shoulder peaks on both higher and lower energy sides were observed. A recent 2D Correlation Spectroscopy study showed that the broad spin-wave band centered at ~ 815 cm$^{-1}$ of hexagonal LuMnO$_3$ single crystal would be contributed by five individual peaks at 741, 783, 812, 839, and 872 cm$^{-1}$ [20]. These indicate that the experimental observed broad spin wave band in hexagonal manganites has complex multiple origins.

A simple method of understanding spin wave is to consider the spin exchange interaction Heisenberg Hamiltonian. Below Néel temperature, the spins of the Mn$^{3+}$ ions forming triangular networks in the a-b plane in hexagonal manganite [9, 21]: each Mn$^{3+}$ ion has two nearest neighbors and four next nearest neighbors. Thus in the a-b plane, there are two exchange interactions J$_1$ (nearest neighbor: intratrimer Mn-Mn interaction) and J$_2$ (next nearest neighbor: intertrimer Mn-Mn interaction), and $H = J_1 \sum_{<i,j>}(S_i \cdot S_j) + J_2 \sum_{<i,k>}(S_i \cdot S_k)$, where $S_i$ is the spin on site $i$, the summation on $j$ is over the nearest-neighbor Mn$^{3+}$ ion pairs and the summation on $k$ is over the next-nearest-neighbor Mn$^{3+}$ ion pairs. Neutron scattering experiments had estimated that J$_1 \approx$ –4.1 meV and J$_2 \approx$ –1.5 meV of hexagonal LuMnO$_3$ single crystal [22]. Then the Heisenberg Hamiltonian model indicates that the broad spin-wave band centered at ~ 815 cm$^{-1}$ would be mainly originated from 4-spin-flip.

For multiple spin flip in the spin-network of magnetic ion with spin state ≥ 1, different types of multiple spin flip states are possible [23]. In hexagonal manganites, the Mn$^{3+}$ ions (spin state of S = 2) forming triangular networks. Then, there would be five possible states of 4-spin-flip, i.e., (1) the 4-spin-flip occurs in one Mn$^{3+}$ ion with all four electrons flipping spin, this is the case proposed in Fig. 3, (2) the 4-spin-flip occurs in two neighboring Mn$^{3+}$ ions with three electrons flipping spin in one ion and another one electron flipping spin in the neighboring ion, (3) the 4-spin-flip occurs in two neighboring Mn$^{3+}$ ions with two electrons flipping spin in both ions, (4) the 4-spin-flip occurs in three neighboring Mn$^{3+}$ ions with two electrons flipping spin in one ion and one electron flipping spin in both the neighboring ions, (5) the 4-spin-flip occurs in four neighboring Mn$^{3+}$ ions with one electron flipping spin in all the four ions. In single crystal, due to the super crystalline quality, all the above five possible states of 4-spin-flip could be observed. These five modes would have similar energy, but different scattering cross section. Thus, the observed 4-spin-flip spin wave scattering of hexagonal LuMnO$_3$ single crystal would be a broad asymmetric band constituted by five individual peaks. This is in good agreement with the spin wave spectrum and 2D Correlation Spectroscopy study. The 2D Correlation Spectroscopy study showed five individual peaks at 741, 783, 812, 839, and 872 cm$^{-1}$ [20], and the observed broad spin wave band also indicated those five individual peaks, as shown in Fig. 4.

Heisenberg Hamiltonian model suggests that the frequency of the five possible states of 4-spin-flip spin wave decreases systematically from (1) to (5). The Raman experiments indicate that mode (3) has the highest scattering intensity. This suggests that the 4-spin-flip occurs in two neighboring $Mn^{3+}$ ions with two spin flip in both ions would have the highest probability. In thin films, this mode would have the major contribution for spin wave scattering. Thus the observed spin wave in thin film has much narrower linewidth and is much more symmetric than the spin wave in single crystal. A systematic comparison studies of pressure dependent Raman, magnetic field dependent Raman, and resonance effect dependent Raman of spin wave scattering between single crystal and thin film would be very helpful for further understanding the physical origins of spin wave in hexagonal manganites.

In conclusion, we presented that the excitation of spin wave in hexagonal manganites can be selectively controlled with on-site Mn *d-d* transition. This property of spin wave in hexagonal manganites would be very helpful for their future applications in magnonic devices. The selective excitation of spin wave also indicated that the spin flip in hexagonal manganites could be selectively controlled, and the processes of spin flip, spin relaxing, and spin wave scattering in hexagonal manganites were proposed. In addition, the physical origin of spin wave correlated with spin flip in hexagonal manganites was discussed.

**Acknowledgments**

X. B. Chen acknowledges the support by the National Natural Science Foundation of China (Grant No. 11574241). I. S. Yang acknowledges the support by the National Research Foundation of Korea (NRF) grant funded by the Korea government (MSIP) (No.

2015001948). X. Wang and S. W. Cheong are supported by the DOE under Grant No. DE-FG02-07ER46382.

**References**

1. J. M. Smith, P. A. Dalgarno, R. J. Warburton, A. O. Govorov, K. Karrai, B. D. Gerardot, and P. M. Petroff, Phys. Rev. Lett. **94**, 197402 (2005).

2. Y. Li, Y. Chye, Y. F. Chiang, K. Pi, W. H. Wang, J. M. Stephens, S. Mack, D. D. Awschalom, and R. K. Kawakami, Phys. Rev. Lett. **100**, 237205 (2008).

3. S. M. Frolov, S. Luscher, W. Yu, Y. Ren, J. A. Folk, and W. Wegscheider, Nature **458**, 868 (2009).

4. M. V. Costache and S. O. Valenzuela, Science **330**, 1645 (2010).

5. D. Serrate, P. Ferriani, Y. Yoshida, S. -W. Hla, M. Menzel, K. Bergmann, S. Heinze, A. Kubetzka, and R. Wiesendanger, Nature Nanotechnology **5**, 350 (2010).

6. V. V. Kruglyak, S. O. Demokritov, and D. Grundler, J. Phys. D: Appl. Phys. **43**, 264001 (2010).

7. P. Rovillain, R. de Sousa, Y. Gallais, A. Sacuto, M. A. Méasson, D. Colson, A. Forget, M. Bibes, A. Barthélémy, and M. Cazayous, Nat. Mater. **9**, 975 (2010).

8. B. Lenk, H. Ulrichs, F. Garbs, M. Münzenberg, Phys. Rep. **507**, 107 (2011).

9. S. Lee, A. Pirogov, M. Kang, K. -H. Jang, M. Yonemura, T. Kamiyama, S. -W. Cheong, F. Gozzo, N. Shin, H. Kimura, Y. Noda and J. -G. Park, Nature **451**, 805 (2008).

10. X. Fabrèges, S. Petit, I. Mirebeau, S. Pailhès, L. Pinsard, A. Forget, M. T. Fernandez-Diaz, and F. Porcher, Phys. Rev. Lett. **103**, 067204 (2009).

11. T. Choi, Y. Horibe, H. T. Yi, Y. J. Choi, W. Wu, and S. -W. Cheong, Nature Materials **9**, 253 (2010).

12. D. Lee, A. Yoon, S.Y. Jang, J.-G. Yoon, J.-S. Chung, M. Kim, J. F. Scott, and T.W. Noh, Phys. Rev. Lett. **107**, 057602 (2011).

13. X. B. Chen, N. T. M. Hien, D. Lee, S. -Y. Jang, T. W. Noh, and I. S. Yang, New J. Phys. **12**, 073046 (2010).

14. X. B. Chen, N. T. M. Hien, D. Lee, S. -Y. Jang, T. W. Noh, and I. S. Yang, Appl. Phys. Lett. **99**, 052506 (2011).

15. X. B. Chen, N. T. M. Hien, K. Han, J. Y. Nam, N. T. Huyen, S. I. Shin, X. Wang, S. W. Cheong, D. Lee, T. W. Noh, N. H. Sung, B. K. Cho, and I. S. Yang, Sci. Rep. **5**, 13366 (2015).

16. A. Muñoz, J. A. Alonso, M. J. Martínez-Lope, M. T. Casáis, J. L. Martínez, and M. T. Fernández-Díaz, Phys. Rev. B **62**, 9498 (2000).

17. J. Park, J. -G. Park, G. S. Jeon, H. -Y. Choi, C. Lee, W. Jo, R. Bewley, K. A. McEwen, and T. G. Perring, Phys. Rev. B **68**, 104426 (2003).

18. A. B. Souchkov, J. R. Simpson, M. Quijada, H. Ishibashi, N. Hur, J. S. Ahn, S. W. Cheong, A. J. Millers, and H. D. Drew, Phys. Rev. Lett. **91**, 027203 (2003).

19. W. S. Choi, S. J. Moon, S. S. A. Seo, D. Lee, J. H. Lee, P. Murugavel, T. W. Noh, Yun Sang Lee, Phys. Rev. B **78**, 054440 (2008).

20. N. T. M. Hien, N. T. Huyen, X. B. Chen, Y. Park, Y. M. Jung, D. Lee, T. W. Noh, S. W. Cheong, and I. S. Yang, J. Mol. Struct., In Press, Available online 17 March 2016.

21. T. J. Sato, S. -H. Lee, T. Katsufuji, M. Masaki, S. Park, J. R. D. Copley, and H. Takagi, Phys. Rev. B **68**, 014432 (2003)**.**

22. H. J. Lewtas, A. T. Boothroyd, M. Rotter, D. Prabhakaran, H. Müller, M. D. Le, B. Roessli, J. Gavilano, P. Bourges, Phys. Rev. B **82**, 184420 (2010)**.**

23. C. Kadolkar, D. K. Ghosh, C. R. Sarma, J. Phys.: Condens. Matter **4**, 9651 (1992).

**Figure Captions**

**Figure 1.** Raman spectra of hexagonal LuMnO$_3$ single crystal at 20 K obtained in the $z(yx)\bar{z}$, $y(zx)\bar{y}$, and $x(yz)\bar{x}$ configurations using 633 nm laser excitation.

**Figure 2.** Raman spectra of hexagonal LuMnO$_3$ single crystal at 20 K obtained in the $z(yx)\bar{z}$ configuration using 633 nm and 514 nm laser excitations. The inset shows that resonance effect of spin wave of hexagonal HoMnO$_3$ thin film [13].

**Figure 3.** Schematic energy diagram of a possible process of spin flip, spin relaxing, and spin wave scattering in hexagonal manganites. In the figure, it is assumed that the 4-spin-flip occurs in one Mn$^{3+}$ ion with all four electrons flipping spin.

**Figure 4.** The broad spin wave band of hexagonal LuMnO$_3$ single crystal is constituted by five individual peaks at 741, 783, 812, 839, and 872 cm$^{-1}$. These peaks are not fitted peaks, they are presented for indicating the possible origins of the broad spin wave band only. The inset shows the 2D Correlation Spectroscopy result [20], which also indicated these five peaks.

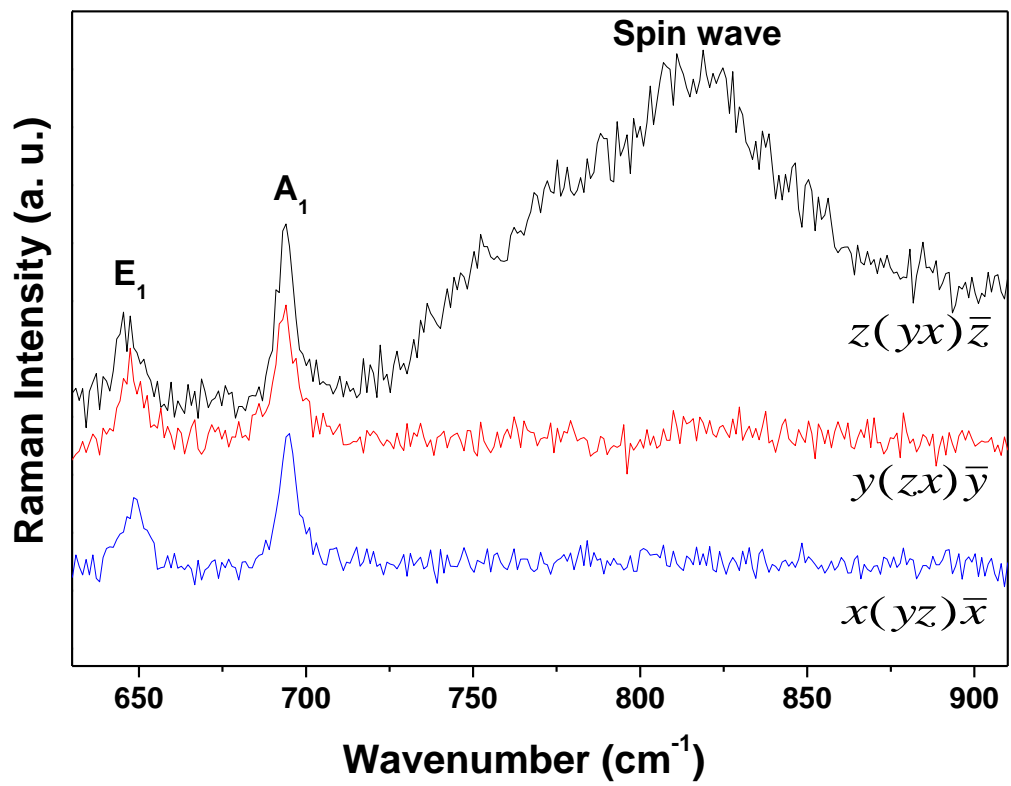

**Figure 1**

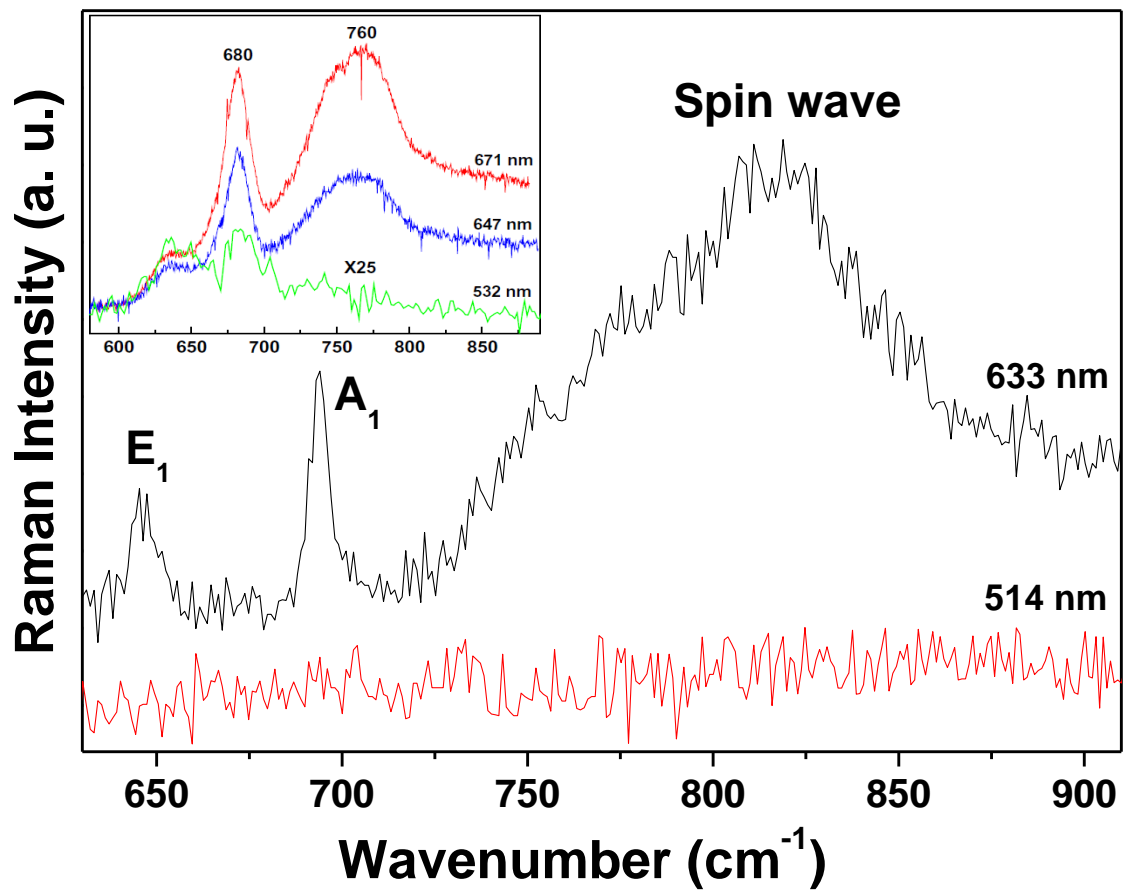

**Figure 2**

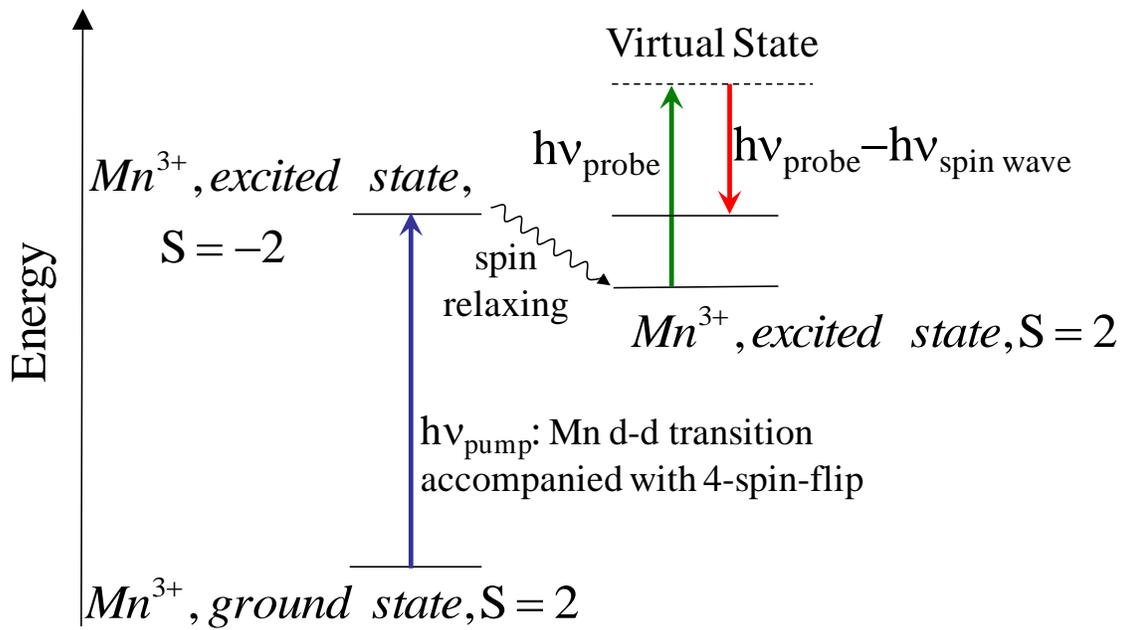

**Figure 3**

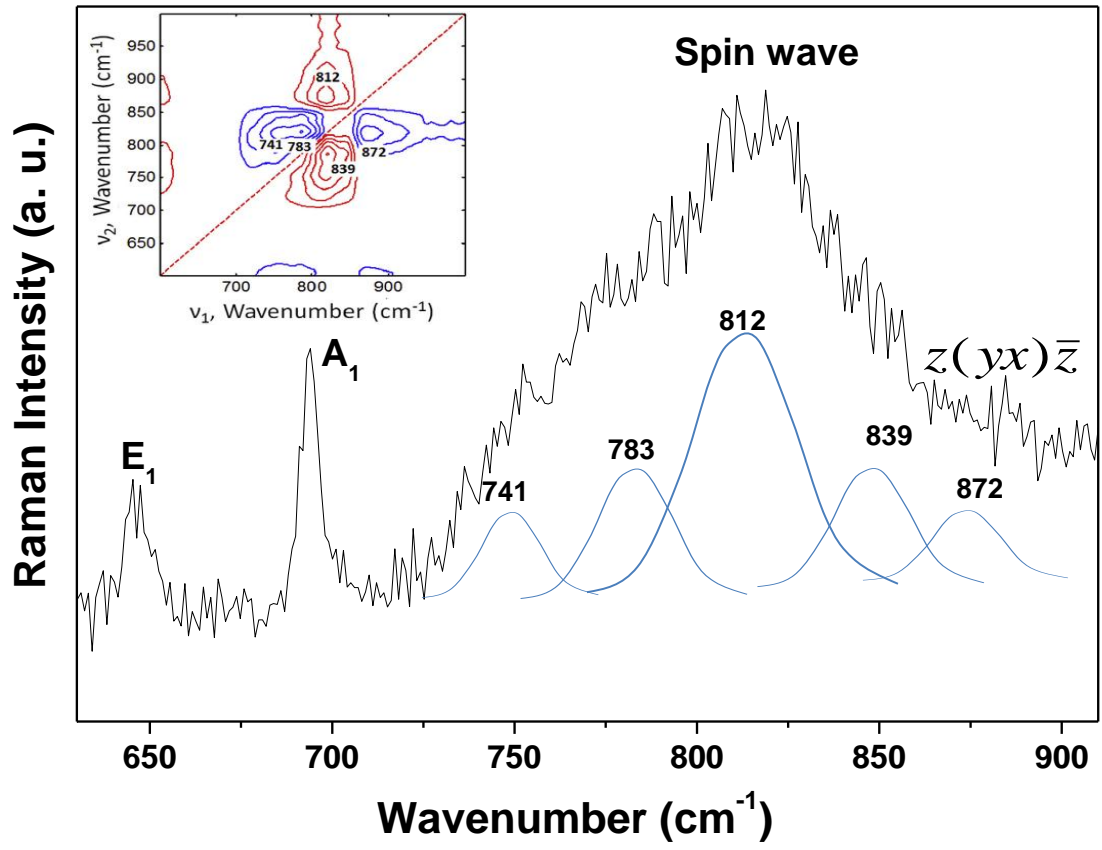

**Figure 4**